*Article*

# An Adaptive Fault-Tolerant Communication Scheme for Body Sensor Networks

Guowei Wu, Jiankang Ren, Feng Xia * and Zichuan Xu

School of Software, Dalian University of Technology, Dalian 116620, China;
E-Mails: wgwdut@dlut.edu.cn (G.W.); rjk.dlut@gmail.com (J.R.); eggerxu@gmail.com (Z.X.)

* Author to whom correspondence should be addressed; E-Mail: f.xia@ieee.org;
  Tel.: +86-411-8757-1582.

**Abstract:** A high degree of reliability for critical data transmission is required in body sensor networks (BSNs). However, BSNs are usually vulnerable to channel impairments due to body fading effect and RF interference, which may potentially cause data transmission to be unreliable. In this paper, an adaptive and flexible fault-tolerant communication scheme for BSNs, namely AFTCS, is proposed. AFTCS adopts a channel bandwidth reservation strategy to provide reliable data transmission when channel impairments occur. In order to fulfill the reliability requirements of critical sensors, fault-tolerant priority and queue are employed to adaptively adjust the channel bandwidth allocation. Simulation results show that AFTCS can alleviate the effect of channel impairments, while yielding lower packet loss rate and latency for critical sensors at runtime.

## 1. Introduction

With recent advances in intelligent (bio-) medical sensors, low-power integrated circuits and wireless networking technologies, Body Sensor Networks (BSNs) have been applied in many areas, especially in human health monitoring [1-5]. By outfitting patients with wireless wearable or implanted vital sign sensors, detailed real-time data on physiological status can be continuously sampled [6,7]. Although BSN shares many of the same challenges with general wireless sensor networks (WSNs), a number of BSN-specific challenges could be specified.

BSNs often demand high degrees of reliability and specific message latency requirements for real-time health monitoring [8]. However, BSNs have fewer and smaller nodes compared with conventional WSNs. Smaller nodes imply smaller batteries, creating stricter constrains on the energy consumed by processing, storage, and communication resources [9]. Usually, BSNs are vulnerable to channel impairments due to body fading effects and/or RF interference [10]. The radio channel assigned to the BSN services is not always clean and sometimes, even the efficient coding schemes used to combat interference may fail. Channel impairments can cause unreliable data transmission and high Bit Error Rate (BER). Hence in some cases the critical data can't be sent to control nodes in time. Consequently, the doctor may make wrong diagnosis and the patient may be delayed to be cured and even die. In addition, the packet loss results in data retransmission. Because the buffer of a biosensor is often very limited, data retransmissions will consume much energy, thus impelling the node to fail. Therefore, it has become crucially important to provide a fault-tolerant communication scheme for BSNs to guarantee reliable data transmission.

In BSNs, biosensors gathering different types of physiological data may have different reliability requirements [11]. For example, heart rate sensors are often considered more important than blood pressure sensors, and hence should be served first under the condition of lack of shared resources (e.g., bandwidth). At the same time, the level of reliability requirements may change dynamically at runtime. For instance, the reliability requirement of the blood pressure sensor might be low when blood pressure readings are in normal range, but the reliability requirement will become much more rigorous when the readings indicate hypotension or hypertension. As a consequence, the system needs to dynamically maintain the reliability requirements of sensors to provide reliability assurance for the sensor nodes with high demand of reliability.

In this paper we present an adaptive and flexible fault-tolerant communication scheme for BSNs, namely AFTCS. When channel impairments occur, AFTCS can provide reliable data transmission for critical sensors by reserving channel bandwidth according to the perceived information about human physiological status, external environment, and the system itself. Fault-tolerant priority and queue are employed to adaptively adjust the channel resource allocation. Simulations have been conducted to evaluate the performance of the proposed scheme. Results are presented and analyzed.

The rest of the paper is organized as follows. In Section 2, we summarize the related work. Section 3 presents the related variables used in the AFTCS scheme and its application scenario. The detailed description of AFTCS is given in Section 4. In Section 5, the performance of AFTCS is evaluated. Finally, we conclude the paper and outline some future work in Section 6.

## 2. Related Work

The growing interest in BSNs and the continual emergence of new techniques have inspired some efforts to study the reliability and quality-of-service (QoS) of BSNs. Otal *et al.* [6,12,13] proposed a novel QoS cross-layer scheduling mechanism based on fuzzy-logic rules for body sensor networks. An energy-saving distributed queuing MAC protocol is adopted. It can guarantee that all packets are served with a specific BER and within particular latency limit while keeping low power consumption.

However, it neglects different reliability requirements of different types of biosensors. In [11], the challenges brought about by BSN applications are presented and a statistical bandwidth strategy, *i.e.*, BodyQoS, is proposed to guarantee reliable data communication. However, the dynamic change in reliability requirements of sensors is not considered in BodyQoS. Natarajan *et al.* highlighted the existence of the inter-user interference effect in BSN from the perspective of network architectures in [14], and make a preliminary investigation of the impact of inter-user interference and implement an instance with a fixed WSN infrastructure to reduce the interference between users in [15]. Braem *et al.* [16,17] modeled probabilistic connectivity in multi-hop body sensor networks to determine ways to improve reliability, which can guarantee k-connectivity between nodes. Qiao *et al.* [18] proposed a multi-homed body sensor network framework and investigate handover strategies during sensor nodes' movement to increase data reliability for BSNs. In [19] a novel packaging technology for BSN based on non-conductive thermoplastic polyurethane adhesive is presented to connect electronic modules with textile circuits in a cost efficient and reliable way.

Although existing schemes [6,11-19] provide some solutions to improve fault-tolerant performance of BSN, designing fault tolerant BSNs to deal with channel impairments is still a challenging issue. In this paper, an adaptive and flexible fault tolerant communication scheme (AFTCS) for BSN is proposed. Major differences between this work and the aforementioned schemes include:

(1) In order to fulfill the reliability requirement of critical sensors, fault-tolerant priority and queue are employed to adaptively adjust the channel resource allocation. Thus it can adaptively provide the reliability assurance for the sensors with high demand of reliability.

(2) A resource reservation method based on dynamic priority queue is presented, including bandwidth measurement, bandwidth requirements calculation and bandwidth allocation methods. In AFTCS, the fault-tolerant priority dynamically changes according to the fault-related information, which can reflect the dynamic change in reliability requirements of sensors. In case of channel impairments, the packet loss rates for critical sensors will be decreased after channel reservation, and the times of retransmission will be reduced, thus lowering the average transmission latency.

## 3. Preliminaries

In this section, we describe the application scenario. The variables used in the AFTCS scheme will be defined. We consider a scenario where the BSN is formed by a collection of biosensors, control nodes and a base station, as shown in Figure 1. In general, the biosensors are wireless wearable or implanted vital sign sensors which consist of a processor, memory, transceiver, sensors and a power unit. Each biosensor node is typically capable of sensing, processing, storing and transmitting the data. The control node periodically sends the sampled data to the medical server in the hospital through the base station, where they are stored for further processing [20].

Due to the size and energy consumption restrictions, biosensors (such as sweat, EKG, temperature) cannot afford heavy computation and communication. Compared with biosensors, the control node (such as a cell phone or PDA) and the base station have comparatively higher transmission rate and computation power. In other words, BSN is a typical asymmetric structure.

We first give the definitions of some variables used in the AFTCS scheme, as listed in Table 1.

**Definition 1:** Fault-tolerant priority $p$. $p$ indicates the level of reliability requirements of a sensor. It can be dynamically adjusted at runtime.

**Definition 2:** Fault-tolerant priority set. Each sensor has a fault-tolerant priority set. Based on the fault-related information, the control node dynamically selects an appropriate priority for the sensor from its priority set.

**Definition 3:** Fault-tolerant priority queue $Q$. The $Q$ is a priority queue managing the fault-tolerant priorities of all the sensors in the system. The sensors in $Q$ are sorted by priority in descending order.

**Figure 1.** BSN application scenario.

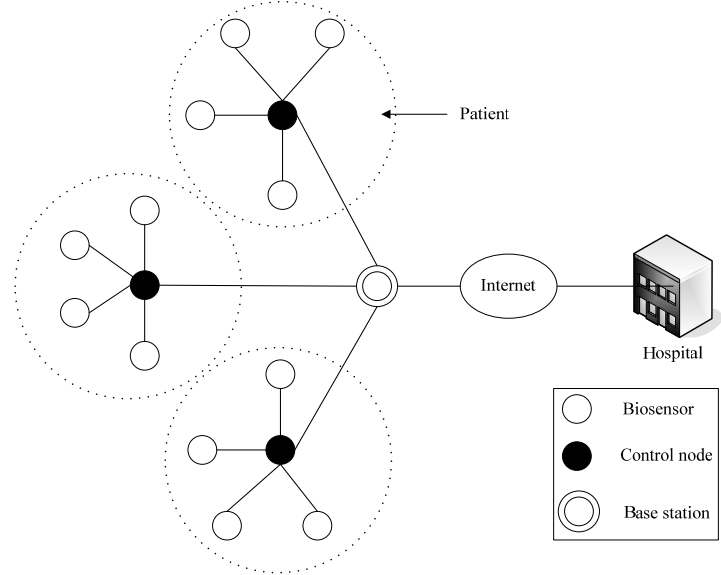

**Table 1.** Variables and Notations.

| Variable | Description |
|---|---|
| $T_{interval}$ | Length of each interval divided in VMAC |
| $N_{pkt}$ | Maximum of packets handled within each interval |
| $S_{pkt}$ | Effective data payload size in bytes |
| $T_{minPkt}$ | Minimum response time for handling a packet request |
| $T_{maxPkt}$ | Maximum response time for handling a packet request |
| $S_i$ | A sensor with a specific function |
| $T_i$ | The time for MAC to send a packet for sensor $S_i$ |
| $D_i$ | Packet number should be sent within $T_i \times D_i$ for sensor $S_i$ |
| $T_{wait}$ | Actual wait time of the control node in a polling process |
| $N_{received}$ | Number of packets received by the control node |
| $Size_{pkt}$ | Data payload of data packet size in bytes |
| $Size_{pollingPkt}$ | Data payload of polling packet size in bytes |
| $BW_{effective}$ | Effective bandwidth |
| $\theta$ | Activation threshold |
| $p$ | Fault-tolerant priority |
| $m_{ip}$ | Priority tuner for fault-tolerant priority $p$ of sensor $S_i$ |
| $\delta_{ip}(t)$ | Adjustment factor of priority $p$ for sensor $S_i$ |
| $\lambda_{ip}(t)$ | The time that $S_i$ wants to maintain $p$ at time $t$ |
| $BW_{ideal}$ | Ideal bandwidth |

**Table 1.** *Cont.*

| | |
|---|---|
| $K_{min}$ | Minimum packets transmitted in each interval for high delay sensitivity sensors |
| $BW_{required}$ | Bandwidth requirements |
| $BW_{CScontrol}$ | Bandwidth reservations for non-polling packets |
| $BW_{SCaware}$ | Bandwidth reservations for fault-related information packets |
| $BW_{SCdata}$ | Bandwidth reservations for sampled data packets |
| $BW_{CSpolling}$ | Bandwidth reservations for polling packets |
| $Q_{reserved}$ | Reservation sensor queue |
| $Q_{removed}$ | The queue containing all sensors removed from reservation |
| $Q_{newreserved}$ | New $Q_{reserved}$ after re-reservation |
| $Q_{newremoved}$ | New $Q_{removed}$ after re-reservation |

## 4. Adaptive Fault-Tolerant Communication Scheme

Figure 2 illustrates the AFTCS scheme. AFTCS consists of three parts: fault-related information collection, fault-tolerant priority queue management and channel resource reservation based on priority queue. Biosensors collect and send physiological data and some fault-related information to the control node. They also execute commands from the control node. The control node analyzes perceived
fault-related information, dynamically changes fault-tolerant priority of the sensors and allocates channel resources to different sensors to reduce the effect of channel impairments. Obviously, the AFTCS scheme is entirely consistent with the asymmetric structure of BSN. In order to guarantee the priorities of critical sensors when the channel resource is scarce, the fault-tolerant priority will be dynamically adjusted. We use the similar resource reservation approach used in [11,21] to address channel impairments. The reserved bandwidth is adjusted according to fault-tolerant priority.

**Figure 2.** Schematic diagram of AFTCS.

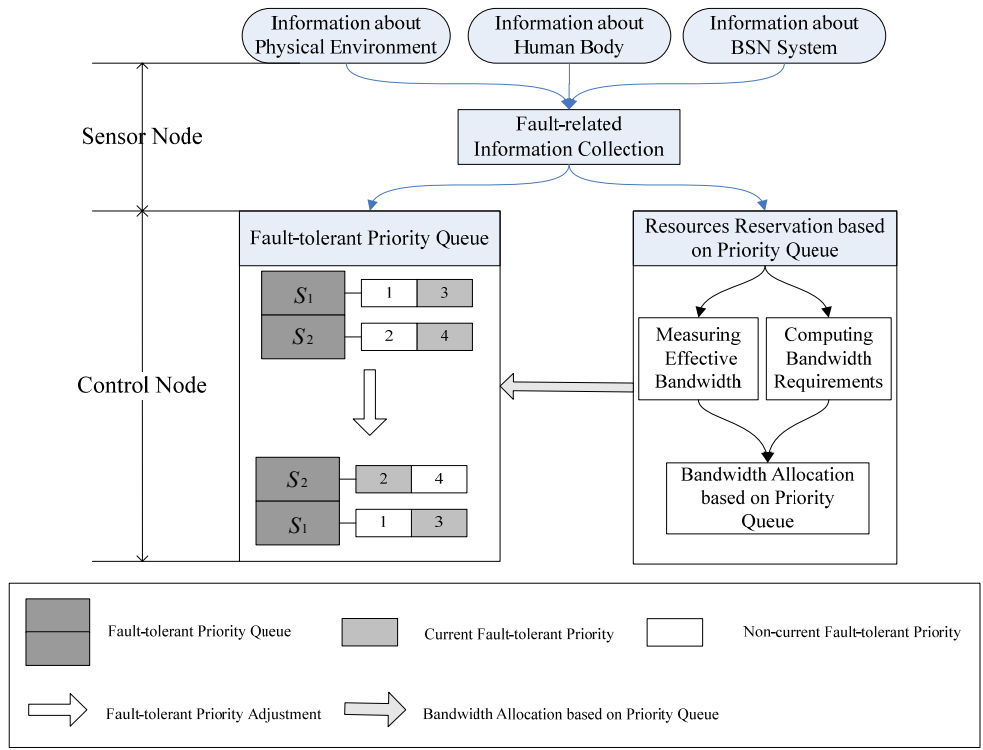

## 4.1. Fault-Related Information Collection

The fault-related information collection module is responsible for obtaining the specific information from human body, physical environment, and the BSN itself. Based on the collected information, we can determine the reliability requirements of the sensors during runtime. In AFTCS, three kinds of fault-related information are collected:

(1) Bioinformation. The bioinformation is mainly physiological data collected by biosensors, such as body temperature, heart rate and blood pressure, *etc.* For a specific biosensor, its reliability requirement might be low when the readings are in normal range (e.g., the range of a body temperature sensor's reading is 36 °C–37.2 °C), but the level of reliability requirements should increase when the readings indicate abnormality (e.g., the reading of a body temperature sensor is higher than 37.5 °C).

(2) Environmental information. The environmental factors (temperature, humidity, light, *etc.*) may affect the health condition of the patient. Therefore, the environmental information will influence the reliability requirement of the biosensor. For example, in warm and humid regions, where water is available as a transmission medium, *Vibrio cholerae* may proliferate rapidly to the level of an infective dose. Hence, a higher level of fault tolerance should be provided for the biosensors that collect *Vibrio cholerae* data under the conditions of high temperature and high humidity.

(3) Runtime system information. These parameters reflect the reliability status of sensor nodes, which include buffer usage, the residual battery life, *etc.* This information can be derived from an inner hardware memory of the biosensor. For example, if the buffer utilization of a certain sensor is larger than 90%, then its reliability requirement will increase; on the other hand, when a sensor's buffer utilization is normal (e.g., 50%), its reliability requirement will not increase or even decrease.

## 4.2. Fault-Tolerant Priority Queue

In order to guarantee the priorities of critical sensors, the control node maintains a fault-tolerant priority queue to manage the current reliability requirements of each sensor. The fault-tolerant priority of a sensor can be statically configured by the clinician through the user interface. It can also be dynamically adjusted according to the perceived fault-related information. An example of the fault-tolerant priority queue containing 5 sensors is shown in Figure 3, where a smaller value indicates a higher priority.

**Figure 3.** A fault-tolerant priority queue.

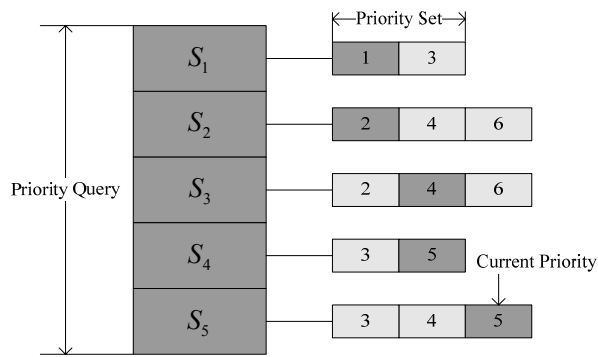

Each sensor has a fault-tolerant priority set. For example, the priority set of sensor $S_1$ is $\{1, 3\}$, *i.e.*, the control node can only choose 1 or 3 as the current fault-tolerant priority for $S_1$. When two sensors have the same fault-tolerant priority, the one requiring less bandwidth reservation (*i.e.*, $BW_{SCdata}$, see Section 4.3) is served first. For example, for $S_4$ and $S_5$ in Figure 3, though their current fault-tolerant priorities are identical, the sensor $S_4$ will be served first because it requires less bandwidth reservation than $S_5$. After the system deployment, each sensor in the system has a default fault-tolerant priority. During the system operation, according to the perceived information, the control node dynamically selects a new fault-tolerant priority as the current priority for the sensor from its fault-tolerant priority set. If the control node perceives channel impairments, it will first provide reliability assurance for the sensors with higher priorities in the queue.

We will give detailed illustration of how to adjust the fault-priority based on some parameters in the following section.

(1). Activation threshold

The parameter $\theta$ is defined as the activation threshold for triggering the adjustment of the fault-tolerant priority of the sensor. For the sake of simplicity, we use one activation threshold for the whole system. As illustrated in Figure 4, the activation threshold is $\theta = 0.4$.

(2). Priority tuner

Priority tuner is used to adjust the current fault-tolerant priority of the sensor. For any fault-tolerant priority $p$ of the sensor $S_i$ in the fault-tolerant priority set, the priority tuner is represented by $m_{ip}$. If $m_{ip}$ exceeds the activation threshold, then the priority $p$ may be activated as the new current priority. All priority tuners of fault-tolerant priorities for the sensor $S_i$ will be set to 0 when its current fault-tolerant priority is reset.

(3). Adjustment factor

$\delta_{ip}(t)$ is defined as the adjustment factor of the fault-tolerant priority $p$ for the sensor $S_i$ at time $t$. If the fault-related information is favorable for the priority $p$, $\delta_{ip}(t)$ will increase. Otherwise, $\delta_{ip}(t)$ will decrease. For example, for the body temperature biosensor $S_{temperature}$, its fault-tolerant priority set is $\{p_1, p_2, p_3\}$, where $p_1 > p_2 > p_3$. According to Table 2, at time $t$, if the body temperature is low without fluctuation, then $\delta_{ip_1}(t) \leq \delta_{ip_3}(t) \leq \delta_{ip_2}(t)$. If the body temperature is high and sharply fluctuating, then $\delta_{ip_1}(t) \leq \delta_{ip_2}(t) \leq \delta_{ip_3}(t)$ and $\delta_{ip_1}(t)$ may be negative.

Table 2. Impact of body temperature on fault-tolerant priority.

| Sensor readings | $p_1$ | $p_2$ | $p_3$ |
|---|---|---|---|
| Low temperature | unfavorable | favorable | favorable |
| Normal temperature | favorable | unfavorable | unfavorable |
| High temperature | unfavorable | favorable | favorable |
| Large fluctuations in temperature | unfavorable | unfavorable | favorable |

(4). Priority acquiescence

The priority acquiescence parameter $\lambda_{ip}(t)$ is used to prevent frequent changes in priority. It represents the time period in which the sensor $S_i$ wants to maintain the activation of the fault-tolerant priority $p$. The function acquiescence$_{ip}(t)$ indicates whether the sensor $S_i$ will acquiesce in its fault-tolerant priority $p$:

$$\text{acquiescence}_{ip}(t) = \begin{cases} 1, & \text{If } p \text{ has been active for } \lambda_{ip}(t) \text{ at time } t \\ 0, & \text{Otherwise} \end{cases} \quad (1)$$

If a fault-tolerant priority $p$ of the sensor $S_i$ has been activated for more than $\lambda_{ip}(t)$ at time $t$, then acquiescence$_{ip}(t)$=1. Otherwise, acquiescence$_{ip}(t) = 0$. This function means that the sensor $S_i$ will acquiesce in the fault-tolerant priority $p$ until $S_i$ has used the priority $p$ for the time period of $\lambda_{ip}(t)$.

The current fault-tolerant priority tuner of the sensor $S_i$ is re-calculated as follows:

$$m_{ip}(0) = 0 \quad (2)$$

$$m_{ip}(t) = [m_{ip}(t-1) + \delta_{ip}(t)] \times \text{acquiescence}_{ip_{current}}(t) \quad (3)$$

**Figure 4.** Switching of fault-tolerant priority.

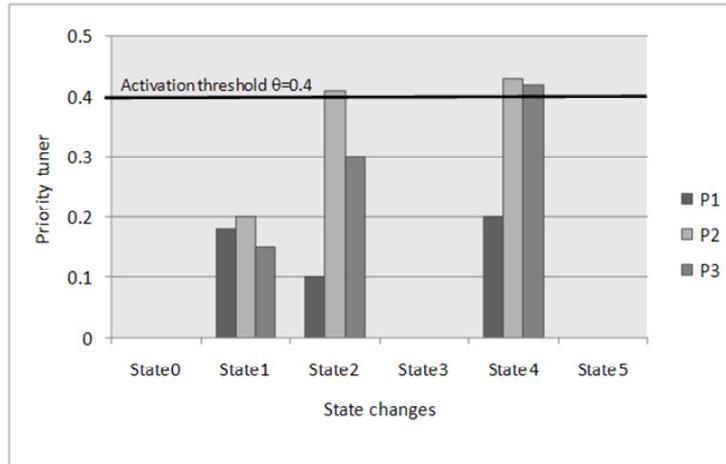

As shown in Figure 4, the current fault-tolerant priority of the sensor $S_i$ is initially $p_{current}$, and $m_{ip}$ is set to 0. If $p_{current}$ is acquiescent at time $t$, then $m_{ip}(t) = 0$ (see e.g., State 0, State 3 and State 5 in Figure 4). Otherwise, $m_{ip}$ will continue to change based on the adjustment factor $\delta_{ip}(t)$ over time unless the priority tuners of some fault-tolerant priorities exceed activation threshold $\theta$, when the control node will activate a new fault-tolerant priority for the sensor $S_i$ (e.g., $p_2$ is activated at State 2 in Figure 4) and all priority tuners of the sensor $S_i$ will be set to 0 (see e.g., State 3 and State 5 in Figure 4). When multiple priorities' tuners exceed activation threshold $\theta$ at the same time, the highest priority will be activated (e.g., $p_3$ is activated at State 4 in Figure 4).

## 4.3. Resource Reservation Based on Priority Queue

In order to reduce the effect of channel impairments, AFTCS adopts a channel bandwidth reservation method, which is similar to the approach used in [11,21]. The control node measures the available channel bandwidth at runtime. When it finds out that the channel impairments occur, it will re-allocate the bandwidth to the sensors according to the fault-tolerant priorities and the bandwidth requirements. The bandwidth reservation method includes three parts: effective bandwidth measurement, bandwidth requirements calculation and bandwidth allocation based on the fault-tolerant priority, as shown in Figure 5. The effective bandwidth represented by the time interval is divided into two parts: reserved bandwidth and best-effort bandwidth. Based on the priority queue, the reserved bandwidth is allocated to the sensors with higher priorities, while the sensors with lower priorities may be served by best-effort communications, and $T_i \times D_i$ represents the time allocated to the sensor $S_i$ to send $D_i$ packets within each time interval. In the cases of initiation of a new sensor, changes in effective bandwidth or sensors' priorities, *etc.* $T_i \times D_i$ may be recalculated to reallocate the effective bandwidth. For example, for $S_3$ and $S_4$, on account of the changes in their priorities, we can reserve bandwidths for $S_4$, while $S_3$ is served by best-effort communications.

**Figure 5.** Bandwidth reservation based on priority queue.

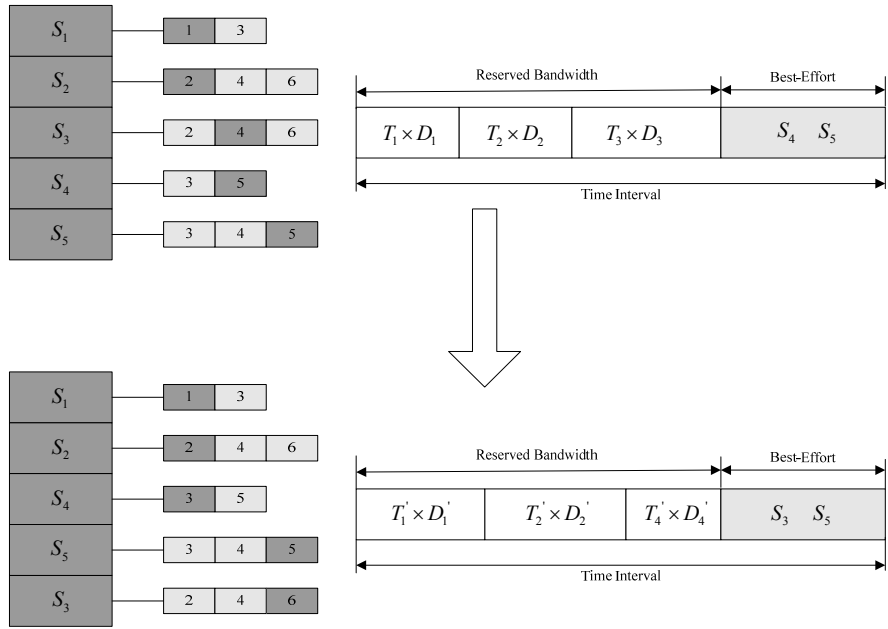

Conventional bandwidth measurement methods usually depend on radio platforms. For example, in CODA [22], each node samples the channel information periodically to get the effective channel bandwidth at runtime. This method is effective for radios with the carrier sense ability. However, it is not suitable for frequency hopping spread spectrum radios such as Bluetooth [23]. AFTCS adopts the same bandwidth measurement method used in [11], which is a radio-agnostic method based on Virtual MAC (VMAC).

The control node sends a polling packet to the biosensor to request the sensor to send acknowledgement packets within specific time. The control node will record the actual waiting time and the number of packets successfully received. In this case the effective channel bandwidth is:

$$BW_{effective} = \frac{Size_{pkt} \times N_{received} + Size_{pollingPkt}}{T_{wait}} \qquad (4)$$

where $Size_{pkt}$ and $Size_{pollingPkt}$ are the data payload of data packet and polling packet size in bytes, respectively. If $T_{wait} = T_i \times D_i + T_{maxPkt}$ and $N_{received} = 0$, then the polling packet is considered lost.

As illustrated in Figure 5, in order to adaptively schedule bandwidth, we should dynamically calculate the values of $D_i$ and $T_i$. In the case of channel impairments, the effective bandwidth $BW_{effective}$ will decrease, and the allocated time for sending one packet $T_i$ will increase, which is given as follows:

$$T_i' = \min\left\{T_{minPkt} \times \frac{BW_{ideal}}{BW_{effective}}, T_{maxPkt}\right\} \qquad (5)$$

where $BW_{ideal} = (N_{pkt} \times S_{pkt} \times 8)/T_{interval}$, $N_{pkt}$ is the maximum number of packets that can be received or transmitted within each $T_{interval}$, assuming a clean channel, $S_{pkt}$ is the data payload of each packet size in bytes, and $T_{maxPkt}$ is the maximum MAC response time for handling a packet transmission request.

In the ideal case, $T_i' = T_{minPkt}$. Hence we can get:

$$D_i' = D_i \times \frac{BW_{ideal} / BW_{effective}}{T_i' / T_{minPkt}} \qquad (6)$$

For highly delay-sensitive sensors, the number of the packets transmitted within each time interval should not be less than $K_{min}$. As a result, the $D_i'$ is recalculated as follows:

$$D_i' = \max\left\{D_i \times \frac{BW_{ideal} / BW_{effective}}{T_i' / T_{minPkt}}, K_{min}\right\} \qquad (7)$$

The method described above assumes that all the sensors have fixed priorities. However, the priorities will be adjusted due to the change of physiological and external environments. Moreover, the bandwidths of some sensors cannot be reserved successfully due to the limited system resources. To tackle these problems, we present a feasible decision-making control of bandwidth allocation.

We compute the requested data bandwidth $BW_{required}$ and compare it with the actual effective bandwidth $BW_{effective}$. Then we allocate the bandwidth based on the dynamic fault-tolerant priorities.

Different from [11], the bandwidth reservation for fault-related information is considered as a portion of $BW_{required}$. As illustrated in Figure 6, the requested data bandwidth $BW_{required}$ includes four parts: (1) Reservations for non-polling packets (e.g., control packets for activation and dormancy) from the control node to sensors $BW_{CScontrol}$; (2) Reservations for fault-related information packets (e.g., the packets including the information of battery usage, buffer usage, *etc.*) from sensors to the control node $BW_{SCaware}$; (3) Reservations for sampled data packets from sensors to the control node $BW_{SCdata}$; (4) Reservations for polling packets from the control node to sensors $BW_{CSpolling}$. We can consequently get the bandwidth requirements of all reservations in the system as follows:

$$BW_{required} = BW_{CScontrol} + BW_{SCaware} + BW_{SCdata} + BW_{CSpolling} \qquad (8)$$

**Figure 6.** Bandwidth reservation requirements.

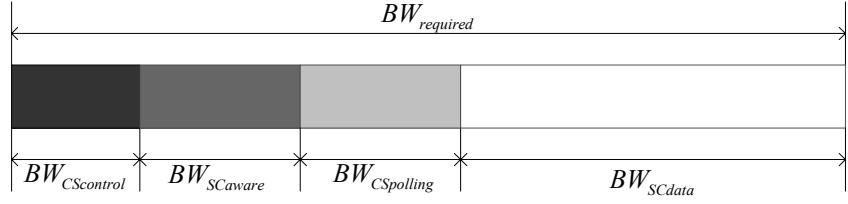

Figure 7 illustrates the process of bandwidth reservation decision-making. $BW_{CScontrol}$ and $BW_{SCaware}$ are reserved fist, and then the bandwidths of $BW_{CSpolling}$ and $BW_{SCdata}$ for sensors are reserved according to the order of fault-tolerant priority queue. We use an approach similar to BodyQoS [11] to divide the effective bandwidth into three parts: $0 \sim B_L \times BW_{effective}$, $B_L \times BW_{effective} \sim B_H \times BW_{effective}$, and $B_H \times BW_{effective} \sim BW_{effective}$, where $0 < B_L < B_H < 1$. The control node allocates unused wireless resources for best-effort communications.

**Figure 7.** Channel reservation decision-making.

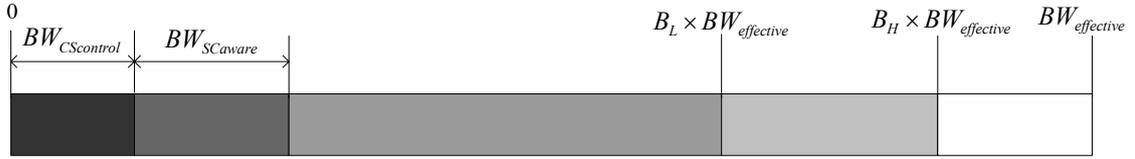

Algorithm 1 presents the pseudocode of the bandwidth reservation control algorithm used in AFTCS. A reservation sensor queue $Q_{reserved}$ and a queue $Q_{removed}$ containing all sensors removed from reservation queue are maintained through minimum heap and maximum heap respectively based on reliability requirements of sensors. If multiple sensors need to re-reserve bandwidth for fault-tolerance simultaneously, the sensors will be processed in priority descending order. The new reservation $S_{add}$ for the sensor is handled according to the total required bandwidth $BW_{required}$ (including the new reservation):

(1) If the total required bandwidth $BW_{required} \leq B_L \times BW_{effective}$, then the new reservation $S_{add}$ is acceptable and $S_{add}$ is added into $Q_{reserved}$.

(2) Under the condition of $B_L \times BW_{effective} < BW_{required} \leq B_H \times BW_{effective}$, if the fault-tolerant priority of $S_{add}$ is not less than the lowest fault-tolerant priority in $Q_{reserved}$, then its new reservation is accepted and $S_{add}$ is added into $Q_{reserved}$. Otherwise, its new reservation is refused and $S_{add}$ is added into $Q_{removed}$.

(3) Finally, in the case of the total required bandwidth $BW_{required} > B_H \times BW_{effective}$, if it can make enough space for the new reservation $S_{add}$ by removing the bandwidth of sensors with lower priorities from $Q_{reserved}$, then the new reservation $S_{add}$ is accepted. Otherwise, it is rejected. In the process of removing the sensors from $Q_{reserved}$, three situations may occur:

- If all sensors with lower priorities than $S_{add}$ in $Q_{reserved}$ have been processed and the aggregate bandwidth of them (*i.e.*, *Sum*) is still less than $BW_{required} - B_H \times BW_{effective}$, then the new reservation of $S_{add}$ is rejected and $S_{add}$ is added into $Q_{removed}$.
- If the aggregate bandwidth (*Sum*) of the sensors ($Q_{temp}$) that may be removed from $Q_{reserved}$ satisfies that $Sum < BW_{required} - B_H \times BW_{effective}$, then the current sensor is added into the temporary queue $Q_{temp}$, and the next sensor with lower priority than $S_{add}$ in $Q_{reserved}$ will be checked.
- If the aggregate bandwidth of sensors ($Q_{temp}$) that may be removed from $Q_{reserved}$ satisfies that $Sum \geq BW_{required} - B_H \times BW_{effective}$, then the new reservation of $S_{add}$ is acceptable; Sensors in $Q_{temp}$ are removed from $Q_{reserved}$ and added into $Q_{removed}$; and $S_{add}$ is added into $Q_{reserved}$.

For the sensors in $Q_{removed}$, they are served by best-effort communications temporarily. However, in the following four situations, their bandwidths may be re-reserved: (1) the fault-tolerant priorities increase; (2) the fault-tolerant priorities of sensors in $Q_{reserved}$ decrease; (3) the bandwidth requirements are reduced; (4) the effective bandwidth increases.

**Algorithm 1.** Bandwidth reservation control algorithm.

---

**Require:** $Q_{reserved}$, $Q_{removed}$, $S_{add}$, $BW_{effective}$, $BW_{required}$, $B_L$, $B_H$
**Ensure:** $Q_{newreserved}$, $Q_{newremoved}$
1:    **if** $BW_{required} \leq B_L \times BW_{effective}$ **then**
2:       $Q_{newreserved} = Q_{reserved}.\text{add}(S_{add})$
3:       $Q_{newremoved} = Q_{removed}$
4:       **return**
5:    **else**
6:       **if** $BW_{required} \leq B_H \times BW_{effective}$ **then**
7:         **if** the priority of $S_{add}$ is not higher than $Q_{reserved}.\text{minPriority}()$ **then**
8:            $Q_{newreserved} = Q_{reserved}$
9:            $Q_{newremoved} = Q_{removed}.\text{add}(S_{add})$
10:           **return**
11:         **else**
12:            $Q_{newreserved} = Q_{reserved}.\text{add}(S_{add})$
13:            $Q_{newremoved} = Q_{removed}$
14:           **return**
15:         **end if**
16:       **else**
17:         $Sum = 0$
18:         $Iterator = Q_{reserved}.\text{begin}()$
19:         **while** $Iterator \neq Q_{reserved}.\text{end}()$ and the priority of $Iterator$ is lower than $S_{add}.\text{priority}()$
20:            $Q_{temp}.\text{add}(Iterator)$
21:            $Sum\mathrel{+}= Iterator.\text{bandwidth}()$
22:            $Iterator\mathord{++}$
23:            **if** $Sum \geq BW_{required} - B_H \times BW_{effective}$ **then**
24:               $Q_{reserved} = Q_{reserved}.\text{remove}(Q_{temp})$
25:               $Q_{newreserved} = Q_{reserved}.\text{add}(S_{add})$
26:               $Q_{newremoved} = Q_{removed}.\text{add}(Q_{temp})$
27:               **return**
28:            **end if**
29:         **end while**
30:       **end if**
31:    **end if**
32:    $Q_{newreserved} = Q_{reserved}$
33:    $Q_{newremoved} = Q_{removed}.\text{add}(S_{add})$

*4.4. Time Complexity Analysis*

**Theorem 1:** The time complexity of AFTCS is $T(n) = O(n\lg n)$, where $n$ is the number of sensors in the system.

**Proof.** The fault-tolerant channel allocation algorithm based on the priorities mainly maintains two fault-tolerant priority queues: the reservation sensor queue $Q_{reserved}$ and the queue $Q_{removed}$ containing all sensors removed from the reservation queue. $Q_{reserved}$ and $Q_{removed}$ are maintained through minimum heap and maximum heap respectively. Suppose the number of sensors in the system is $n$ and the number of sensors in $Q_{reserved}$ is $m$ ($0 \leq m \leq n$), then the number of sensors in $Q_{removed}$ is $n-m$. For $Q_{reserved}$, the time complexity for building the minimum heap is $O(m)$; the time complexity for the insertion or removal of the sensor is $O(\lg m)$; the time complexity for the traversal of all sensors is $O(m\lg m)$. Similarly, for $Q_{removed}$, the time complexity for building the maximum heap is $O(n-m)$; the time complexity for the insertion or removal of the sensor is $O(\lg(n-m))$; the time complexity for the traversal of all sensors is $O((n-m)\lg(n-m))$. Therefore, the time complexity is as follows.

$$T(n) = O(m\lg m) + O((n-m)\lg(n-m)) = O(n\lg n) \qquad (9)$$

## 5. Performance Evaluation

We evaluated AFTCS on the Castalia simulator [24]. Castalia is an open source, discrete event-driven simulator based on OMNeT++ [25]. The experiment simulates a typical body sensor network, in which sensors measure a person's physiological parameters. We configure the body sensor network with three types of biosensors: ECG sensor, $SpO_2$ sensor and Temperature sensor. All nodes adopt Castalia standard CC2420 IEEE802.15.4 radios. In the whole experiment seven fault-tolerant priorities are adopted. The range of fault-tolerant priority is 0~6, where a smaller value indicates a higher priority. Tables 3 and 4 describe the detailed simulation parameters and sensor node specifications, respectively. We compare our AFTCS scheme with the BodyQoS scheme [11].

Table 3. Simulation settings.

| Parameter | Value |
| --- | --- |
| Number of sensor types | 3 |
| Wireless channel model | Log shadowing wireless model |
| Path loss exponent | 2.4 |
| Collision model | Additive interference model |
| Physical and MAC layer | IEEE 802.15.4 standard |
| Data transmission rate | 250 Kbps |
| Buffer size | 1,024 KBytes |
| Max physical layer frame size | 127 Bytes |
| Physical layer frame overhead | 6 Bytes |
| MAC layer frame overhead | 13 Bytes |
| Ideal noise floor | −100 dBm |
| Simulation time | 600 s |
| Number of simulation runs | 50 |

The simulation includes six time periods. In the first time period (0 s~100 s), the system operates in an ideal state, and the readings of all sensors are normal. In the second time period (100 s~200 s), noise signals are generated to introduce the channel impairment. The noise floor is increased to −80 dBm. In the third time period (200 s~300 s), the noise floor is increased to −70 dBm. Besides, the readings of the temperature sensor become high with large fluctuations. In the fourth time period (300 s~400 s), the readings of the SpO₂ sensor become exceptional and the readings of the temperature sensor become normal. In the fifth time period (400 s~500 s), the noise floor is reduced to −80 dBm. In the sixth time period (500 s~600 s), the noise floor becomes ideal. In other words, the system operates in an ideal state once again.

Table 4. Sensor node specifications.

| Parameter | ECG | SpO$_2$ | Temperature |
| --- | --- | --- | --- |
| Payload size (Bytes) | 50 | 25 | 2 |
| Transmission rate (packets/s) | 10 | 8 | 1 |
| High delay sensitivity | Yes | Yes | No |
| Priority set | {0, 1, 2, 3} | {2, 3, 4, 5} | {3, 4, 5, 6} |
| Initial priority | 2 | 4 | 5 |

In the whole experiment, two metrics are used for performance evaluation: packet loss rate and average packet latency. The packet loss rate is used to evaluate the performance of reliability and the average packet latency is used to evaluate the performance of timeliness.

Figure 8 presents the fault-tolerant priority changes of each sensor during the simulation. In the first time period, the fault-tolerant priorities of all sensors decrease, because they have operated in the fault-free state for a certain time. In the second time period, the priorities of the ECG sensor and the SpO₂ sensor increase due to the channel impairment, while the priority of the temperature sensor doesn't change, because its channel sensitivity is relatively low. In the third time period, the priorities of the ECG sensor and the SpO₂ sensor further increase due to the increase of the noise level. The priority of the temperature sensor increases rapidly because of the abnormality of readings and the channel impairment. In the fourth time period, the priority of the SpO₂ sensor increases, and the priority of the temperature sensor decreases for the changes in the readings. In the fifth time period, there is a decrease in the priority of the ECG sensor owing to the reduction of the noise level. In the sixth time period, the fault-tolerant priorities of all sensors decrease because they have operated in the fault-free state for a certain time. Obviously, the reliability requirements of different biosensors can be maintained at runtime by fault-tolerant priority.

**Figure 8.** Fault-tolerant priority changes.

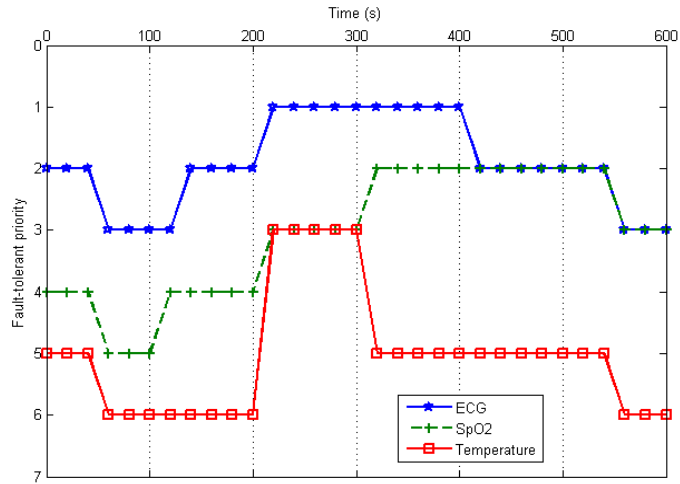

The packet loss rates of BodyQoS and AFTCS are shown in Figure 9. In time Periods 1–2, the packet loss rates of BodyQoS and AFTCS are basically the same. However, from Period 3, for the sensors with higher demand of reliability, AFTCS achieves lower packet loss rates than BodyQoS. The reason is that, AFTCS re-allocates bandwidth according to the new fault-tolerant priorities to first provide reliability assurance for the sensor nodes with higher demand of reliability, while BodyQoS still allocates bandwidth according to the original fixed priorities. It is worth noting that during time periods 3, 5 and 6, two sensors have the same fault-tolerant priorities, and the one requiring less channel resources is served first.

**Figure 9.** Packet loss rate.

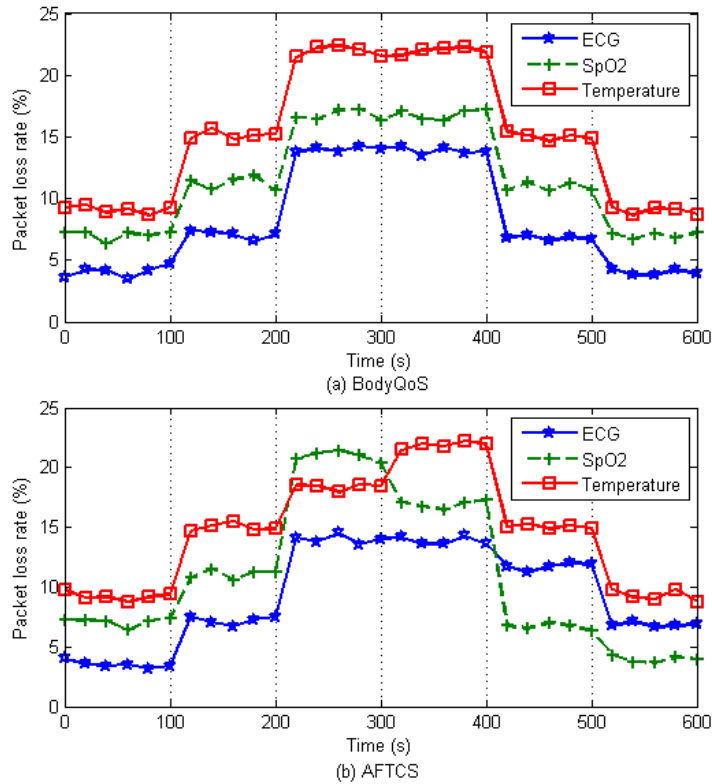

As can be seen from Figure 10, for the sensors with higher demand of reliability, AFTCS achieves lower average packet latency than BodyQoS at runtime. With AFTCS, the transmission of fault-related packets has little effect on packet latency. This is because the fault-related packets are transmitted only when some failures occurred or may occur. Taking the fault-tolerant data packets of the buffer overflow failure for example, the sensors don't send the buffer-overflow-related information to the control node unless the buffer utilization is higher than 80% or 90%. Therefore, basically it doesn't affect the reliable data communication of BSN.

**Figure 10.** Average packet latency.

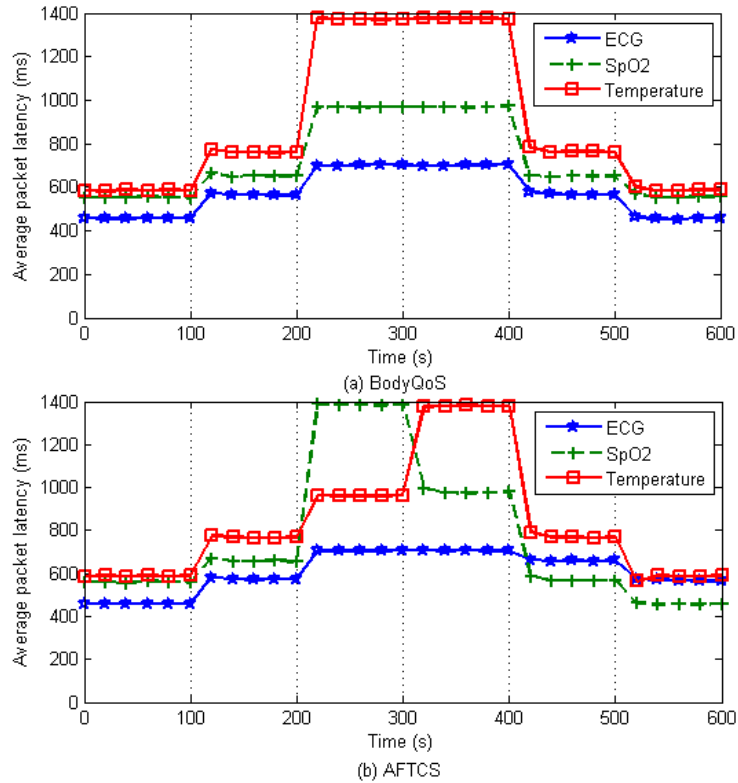

## 6. Conclusions

In this paper, we have presented an adaptive fault-tolerant communication scheme (AFTCS) based on fault-tolerant priorities for BSNs. AFTCS can tolerate channel impairments by exploiting resource reservation. The fault-tolerant priority and queue are employed to dynamically adjust the channel bandwidth allocation so as to fulfill the reliability requirements of critical sensors. Simulation results show that AFTCS can reduce the effect of channel impairments while guaranteeing lower packet loss rates and latency for the sensors with higher demand of reliability at runtime. The primary contributions of this paper are summarized as follows:

(1) An asymmetric fault-tolerant architecture is proposed, in which resource-constrained sensor nodes do little processing and the control node with abundant resources performs the majority of fault-tolerant operations.

(2) An adaptive priority management method is presented, which can dynamically adjusts the fault-tolerant priorities of sensors according to the perceived information, thus guaranteeing the priorities of critical sensors during runtime.

(3) A resource reservation method based on dynamic priority queue is presented. In case of channel impairments, the packet loss rates for critical sensors will be decreased after channel reservation, and the times of retransmission will be reduced, thus shortening the average transmission latency.

In the future, we will design a parameter update strategy to configure parameters dynamically during runtime based on knowledge learned from previous experiences and evaluate AFTCS (e.g., in terms of energy efficiency) by means of more extensive simulations. We also consider implementing and testing the scheme on real-life BSNs that can take advantage of the adaptive and flexible fault-tolerant communication enabled by AFTCS.

## Acknowledgements

This work was partially supported by the Natural Science Foundation of China under Grants No. 60703101 and No. 60903153, and the Fundamental Research Funds for the Central Universities.